\documentclass[prb,twocolumn,showpacs,preprintnumbers,amsmath,amssymb,superscriptaddress]{revtex4}

\usepackage{graphicx}
\usepackage{dcolumn}
\usepackage{bm}

\begin{document}

\preprint{APS/123-QED}

\title{The Effects of Stoichiometry and Substitution in quasi-one-dimensional iron-chalcogenide BaFe$_{2}$S$_{3}$}

\author{Yasuyuki Hirata}
\affiliation{
Institute for Solid State Physics, The University of Tokyo, Kashiwa, Chiba 277-8581, Japan
}

\author{Sachiko Maki}
\author{Jun-ichi Yamaura}
\affiliation{
Materials Research Center for Element Strategy, Tokyo Institute of Technology, Yokohama, Kanagawa 226-8503, Japan
}

\author{Touru Yamauchi}
\affiliation{
Institute for Solid State Physics, The University of Tokyo, Kashiwa, Chiba 277-8581, Japan
}

\author{Kenya Ohgushi}
\affiliation{
Institute for Solid State Physics, The University of Tokyo, Kashiwa, Chiba 277-8581, Japan
}
\affiliation{
Department of Physics, Tohoku University, Sendai, Miyagi 980-8578, Japan
}


\begin{abstract}
The effects of off-stoichiometry and elemental substitution on electronic properties of iron-based ladder compound BaFe$_{2}$S$_{3}$ are investigated. Resistivity and magnetization are revealed to be  quite sensitive to the stoichiometry of Fe atoms, and 10 \% deficiency at Fe sites reduces the antiferromagnetic transition temperature by 40 K. The antiferromagnetic transition temperature decreases even faster and collapse to zero with hole doping through 10 \% K substitution at Ba site, while the antiferromagnetic ordering phase remains with electron doping through 20 \% Co substitution at Fe site. Such electron-hole asymmetry is opposite to two-dimensional iron-based superconductors, and can be explained on the basis of both itinerant and localized electronic pictures.

\end{abstract}

\pacs{74.70.Xa, 72.80.Ga, 75.50.Ee}

\maketitle

\section{Introduction}

Since the superconductivity in LaFeAsO$_{1-x}$F$_{x}$ has been discovered,\cite{kamihara} extensive studies have been conducted on iron-based compounds, which is now a large family of high-temperature superconductors aside from the copper oxides. In contrast to the single-orbital Mott physics in copper oxides, the physics underlying in iron pnictides and chalcogenides is more complicated owing to the multi-orbital character. For example,  stripe-type antiferromagnetic structure is considered to originate not only from the Fermi-surface nesting but also from the orbital ordering, which appears at the almost same critical temperature with that of the magnetic ordering. Consequently, as possible candidates of the paring glue of Cooper pairs, magnetic fluctuations as well as orbital fluctuations are theoretically argued.\cite{mazin,kuroki,kontani} To understand the mechanism of iron-based high-temperature superconductivity, considerable experimental efforts to elucidate the role of each characteristic behaviors of charge, spin, and orbital in related materials is necessary.

\begin{figure}
\includegraphics[width=8cm]{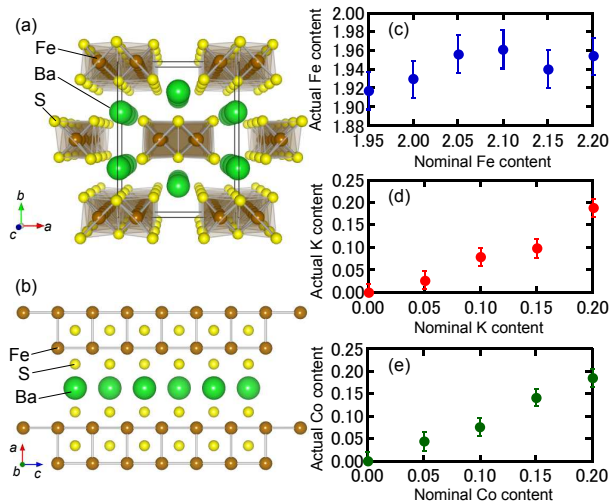}%
\caption{\label{fig:1} (a), (b) The crystal structure of BaFe$_{2}$S$_{3}$. (c-e) The relations between nominal content and actual content estimated by SEM/EDX analysis of (c) Fe in BaFe$_{2+\delta}$S$_{3}$, (d) K in Ba$_{1-x}$K$_{x}$Fe$_{2}$S$_{3}$, and (e) Co in BaFe$_{2-y}$Co$_{y}$S$_{3}$.}
\end{figure}

Controlling the dimensionality is an effective strategy for the study of electronic structures. Similar to the existence of chain-type SrCuO$_{2}$, Sr$_{2}$CuO$_{3}$,\cite{hiroi} and ladder-type Sr$_{14}$Cu$_{24}$O$_{41}$ in copper oxides,\cite{maccarron} there are iron-based materials with quasi-one-dimensional structures. Characteristic examples are iron chalcogenides with a ladder structure $A$Fe$_{2}X_{3}$ ($A$ = K, Rb, Cs, and Ba; $X$ = S, Se, and Te),\cite{nambu,du,dagotto,du_co,dong,takahashi,arita,yamauchi} of which the crystal structure is shown in  Fig.~\ref{fig:1} (a) and (b).\cite{hong,klepp} Even though the local structure of Fe atoms, which are coordinated tetrahedrally by ligand $X$ atoms, resembles that in two-dimensional (2D) iron-based superconductors, one-dimensionality of the two-leg-ladder configuration in $A$Fe$_{2}X_{3}$ results in quite distinct electric conductions. In contrast to the metallic conduction in most 2D iron-based superconductors, all of $A$Fe$_{2}X_{3}$ known so far are insulators under ambient pressure. A particularly important point is that $A$Fe$_{2}$Se$_{3}$ ($A$ = K and Cs), in which the average Fe valence is mixed-valent 2.5+, has even higher resistivity than BaFe$_{2}$Se$_{3}$ with the average Fe valence of 2+, despite the absence of charge ordering.\cite{du} On the other hand, even though the distinct electrical conduction, the magnetism of $A$Fe$_{2}X_{3}$ has a close resemblance to that of the 2D iron-based superconductors: BaFe$_{2}$Se$_{3}$ shows 2 $\times$ 2 block-type magnetic structure\cite{nambu} just like that of $A_{0.8}$Fe$_{1.6}$Se$_{2}$ ($A$ = K and Rb);\cite{ye,mwang} BaFe$_{2}$S$_{3}$ shows stripe-type magnetic structure with the spin direction along the rung direction,\cite{takahashi} which is similar to that of LaFeAsO$_{0.5}$H$_{0.5}$;\cite{hiraishi} and $A$Fe$_{2}$Se$_{3}$ ($A$ = K, Rb, and Cs) show stripe-type magnetic structure with the spin direction along the leg direction,\cite{du} which resembles that of LaFeAsO and $A$Fe$_{2}$As$_{2}$ ($A$ = Sr, and Ba).\cite{qureshi,huang} Note that the ordered magnetic moments in the ladder compounds are 1.2-2.8 $\mu_{\rm B}$, and are much reduced from the localized moments expected for the high-spin state of Fe$^{2+}$ ions, 4 $\mu_{\rm B}$. All of these observations evoke fundamental questions whether the itinerant or localized picture is correct to describe the magnetism of the iron-based materials.\cite{hansmann}

In the light of the pressure-induced superconductivity in Sr$_{14}$Cu$_{24}$O$_{41}$,\cite{uehara} iron-based ladder compounds have been considered to be possible candidates for new superconductors. Indeed, Takahashi {\it et al.} recently discovered that BaFe$_{2}$S$_{3}$ shows superconductivity with the critical temperature ($T_{c}$) of 14 K under the pressure of 11 GPa.\cite{takahashi} It turned out that the appearance of superconductivity is quite sensitive to the nominal composition  and the growth condition, and that the superconducting signal is maximally enhanced at slightly off-stoichiometric BaFe$_{2.1}$S$_{3}$ in nominal composition. Interestingly, the antiferromagnetic transition temperature at the ambient pressure ($T_{\rm N}$) is highest at this composition.\cite{yamauchi} These observations indicate that the effect of off-stoichiometry is very important for understanding electronic properties. Moreover, considering that the emergence of superconductivity on carrier doping by elemental substitutions in 2D iron-based superconductors, one can expect that the carrier doping sensitively changes $T_{\rm N}$ as well as $T_{c}$ in BaFe$_{2}$S$_{3}$. Because the compound is a Mott insulator, carrier doping effect is crucial even in the normal state under ambient pressure.

In this paper, we report the effects of off-stoichiometry and carrier doping on the electronic and magnetic properties of BaFe$_{2}$S$_{3}$. It turned out that both the electric conduction and magnetic properties are very sensitive to off-stoichiometry probably owing to quasi-one-dimensionality, and it is suggested that actual stoichiometric sample BaFe$_{2}$S$_{3}$ is grown from nominally BaFe$_{2.1}$S$_{3}$ composition. The phase diagram of K and Co substituted compounds shows a dome-like antiferromagnetic phase as a function of compositions with electron-hole asymmetry opposite to 2D iron-based superconductors. We discuss these phenomena on the basis of both the itinerant and localized pictures.

\section{Experiment}

Single crystals of BaFe$_{2+\delta}$S$_{3}$, Ba$_{1-x}$K$_{x}$Fe$_{2}$S$_{3}$, and BaFe$_{2-y}$Co$_{y}$S$_{3}$ were grown by the melt-growth method. Commercially available K$_{2}$S, BaS, Fe, Co, and S powders with a total amount of 1.5 g were mixed in the target molar ratio within a glove-box filled with nitrogen gas. The mixture was put into a carbon crucible, and the crucible was sealed in a quartz tube with Ar gas of 0.3 atm. The quartz tube was heated up to 1100 $^{\circ}$C, kept for 24 hours, and then cooled to 750 $^{\circ}$C for 24 hours. The chemical composition of the products was analyzed by the energy dispersive X-ray spectrometer equipped with the scanning electron microscopy (SEM/EDX). Powder X-ray diffraction experiments were done by X-ray diffractometer (Rigaku, SmartLab) using a Cu-K$\alpha$ radiation. Single-crystal X-ray diffraction experiments were carried out on a curved imaging plate (Rigaku, R-AXIS RAPID-II) using a Mo-K$\alpha$ radiation (Rigaku, VariMax). Resistivity was determined by using a commercial setup (Quantum Design, PPMS) with a use of conventional four probe method. Magnetic susceptibility was measured with a superconducting quantum interference device (SQUID) magnetometer (Quantum Design, MPMS).

\section{The Effect of Stoichiometry}

\begin{figure}
\includegraphics[width=7cm]{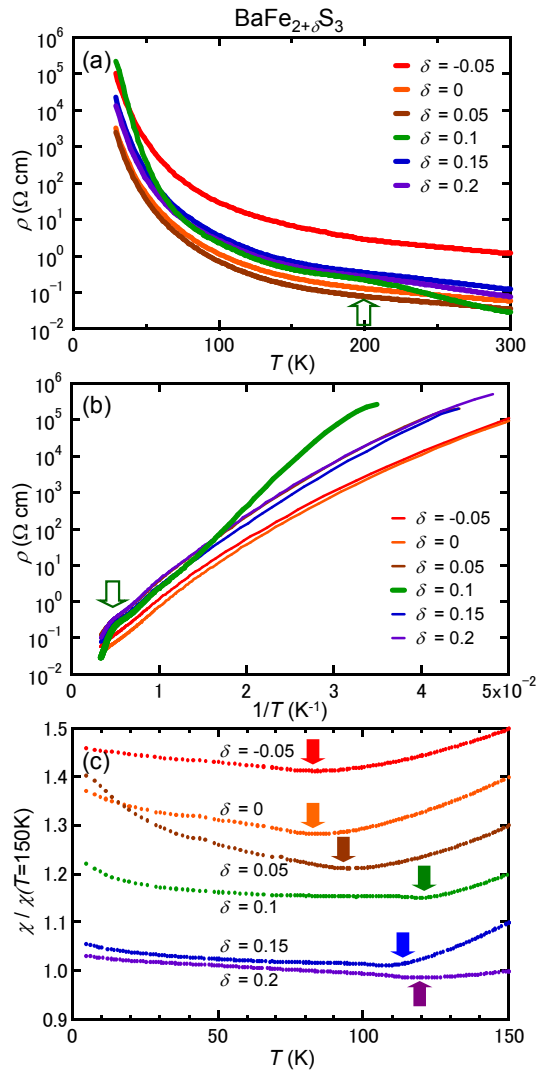}%
\caption{\label{fig:2} (a) Temperature ($T$)-dependence of resistivity ($\rho$) of BaFe$_{2+\delta}$S$_{3}$. The arrow indicates resistivity anormaly which is most distinct in $\delta$ = 0.1 composition. (b) Arrhenius plot of $\rho$ of BaFe$_{2+\delta}$S$_{3}$. (c) Normalized magnetic susceptibility under the magnetic field of 5 T $\chi$/$\chi(T = 150 {\rm K})$ for BaFe$_{2+\delta}$S$_{3}$. Each curve is shifted by the offset of 0.1. The arrows indicate the antiferromagnetic transition temperature ($T_{\rm N}$).}
\end{figure}

Figure~\ref{fig:1} (c) shows the relation between nominal and actual Fe content, $2+\delta$, in BaFe$_{2+\delta}$S$_{3}$. We here note that the sample is made from starting materials of ${\rm BaS} : {\rm Fe} : {\rm S} = 1 : 2+\delta : 2$. The actual Fe content is estimated by taking the molar ratio between the number of Fe atoms and the total number of atoms, which is evaluated by the SEM/EDX analysis. With increasing the nominal Fe content, $2+\delta$, the actual Fe content slowly but monotonically increases in the small nominal Fe content region $\delta < 0.1$; however, there is no further increase in the large Fe content region $\delta > 0.1$, keeping the actual Fe content smaller than 2. These observations indicate that there is deficiency in Fe sites in spite of the excess Fe ratio in the nominal compositions. For the simplicity, we hereafter use the nominal Fe content to distinguish the samples. We stress that even though $\delta > 0$ the Fe sites are actually deficient. Note that the powder X-ray diffraction patterns of BaFe$_{2+\delta}$S$_{3}$ (data not shown) show no systematic change against $\delta$, which indicates that they have a same crystal structure within the experimental accuracy and Fe deficiency is small.

Figure \ref{fig:2} (a) shows the temperature ($T$)-dependence of the resistivity ($\rho$) of BaFe$_{2+\delta}$S$_{3}$ along the $c$-axis (parallel to the ladder). For all $\delta$, $\rho$ shows an insulating behavior, which is due to the strong electron correlation effect in low-dimensional structure. However, if one takes a closer look at the low-temperature regime, one can notice that $\rho$ of $\delta$ = 0.1 compound differs from the other compounds. As clearly shown in Fig.~\ref{fig:2} (b), the plot of $\rho$ in the logalithmic scale against the inverse of the temperature 1/$T$ (Arrhenius plot) obeys a linear relationship only for $\delta$ = 0.1 compound; this is the thermal-activation-type conduction. The gap energy estimated from the line slope is 47 meV. On the other hand, the plots for $\delta \neq$ 0.1 compounds show a substantial deviation from the linear relationship in a concave-down manner. Instead, one can see a linear relationship if one plot $\rho$ in the logalithmic scale against 1/$T^{1/2}$ (data not shown); this indicates the one-dimensional variable range hopping (1D VRH) conduction. Such $\delta$-dependencies of $\rho$ together with the relationship between the nominal and actual Fe contents (Fig. 1 (c)) strongly suggests that the true stoichiometric compound BaFe$_{2}$S$_{3}$ is grown from the nominally $\delta$ = 0.1 composition. Actually, our x-ray single-crystal structural analysis revealed the stoichiometric chemical composition in it; the structural parameters at the room temperature is summarized in Tab.~\ref{tab:1}. In other compounds, it is likely that the off-stoichiometry, which should be as small as $\sim$ 0.02, induces a small number of localized carriers into ladders and/or impurity sites sitting between the energy gap, which results in the variable range hopping conduction.

\begin{table}
\caption{\label{tab:1} 
Fractional atomic coordinates and equivalent displacement parameters $U_{eq}$ at the room temperature. The nominal composition is BaFe$_{2.1}$S$_{3}$ (see the text for detail). The compounds crystallizes with space group $Cmcm$ (No. 63). The lattice constants are $a$ = 0.87781(15) nm, $b$ = 1.1233(2) nm, and $c$ = 0.52884(7) nm. The reliability indices of this fit are $R$1 = 3.47 \% and w$R$ = 7.33 \%. For each site, no trace of atomic deficiency is detected.}
\begin{tabular}{cclllll}
\hline
 & Wycoff & $x$ & $y$ & $z$ & $U_{eq}$ ($10^{-4}$nm$^{2}$) \\ \hline
Ba & 4c & 0.5 & 0.18635(4) & 0.25 & 2.29(2) \\
Fe & 8e & 0.15370(7) & 0 & 0 & 1.57(2) \\
S1 & 4c & 0 & 0.11584(16) & 0.25 & 1.61(3) \\
S2 & 8g & 0.29235(15) & 0.12182(15) & 0.25 & 2.37(3) \\ \hline
\end{tabular} 
\end{table}

Next, we move to  $\rho$ behavior in the high temperature region.  We notice that a kink feature is present at around $T$ = 200 K at $\delta$ = 0.1 (indicated by the arrow in  Fig. 2 (a)). This  feature is maximally clear in the curve of $\delta$ = 0.1, however, as shown in Fig.~2 (b), all curves exhibits a similar feature. The origin of this anomaly is not unraveled yet; a possible candidate is a structural transition which involves the orbital ordering. The orbital ordering is commonly seen in the 2D iron-based superconductors.\cite{huang,lee,chuang,nakayama} However, our preliminary x-ray diffraction experiments at 167 K could not detect any superlattice reflections. We therefore conclude that, if any, the structure displacement is tiny across the structural transition.

\begin{figure}
\includegraphics[width=5cm]{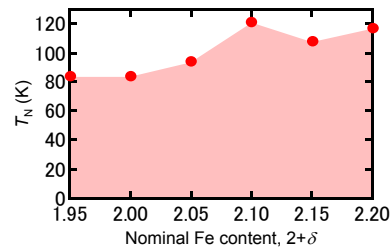}%
\caption{\label{fig:3} The relation between the antiferromagnetic transition temperature ($T_{\rm N}$) and the nominal Fe content of BaFe$_{2+\delta}$S$_{3}$.}
\end{figure}

The $T$-dependence of the magnetic susceptibility ($\chi$) of BaFe$_{2+\delta}$S$_{3}$ under the magnetic field of 5 T parallel to the $c$-axis (the ladder direction) is shown in Fig.~\ref{fig:2} (b). A small dip feature in $\chi$ curve shown by an arrow indicates the antiferromagnetic transition. The N\'{e}el temperature $T_{\rm N}$ is plotted against the nominal Fe content $2+\delta$ in Fig.~\ref{fig:3}. The $T_{\rm N}$ value reaches the highest value 122 K at $\delta$ = 0.1, which is another evidence that the true stoichiometric composition is realized in the nominal $\delta$ = 0.1 composition. With decreasing $\delta$ from 0.1,  $T_{\rm N}$ gradually decreases down to 82 K at $\delta$ = 0. Since Fe deficiency effectively works as hole-doping, one can interpret the reduction of $T_{\rm N}$ in terms of the weakened nesting properties; another simpler interpretation is that Fe deficiency just works as an impurity, hindering the system from the long-range magnetic ordering.  For $\delta >$ 0.1, one can see no meaningful variation in $T_{\rm N}$, hinting at that  neither excess Fe atom nor Ba/S deficiency is possible in this compound. This is consistent with small variation of actual Fe content as a function of nominal Fe content at $\delta >$ 0.10, as shown in Fig. 1 (c) 

\section{The Effect of Substitution}
\begin{figure}
\includegraphics[width=7cm]{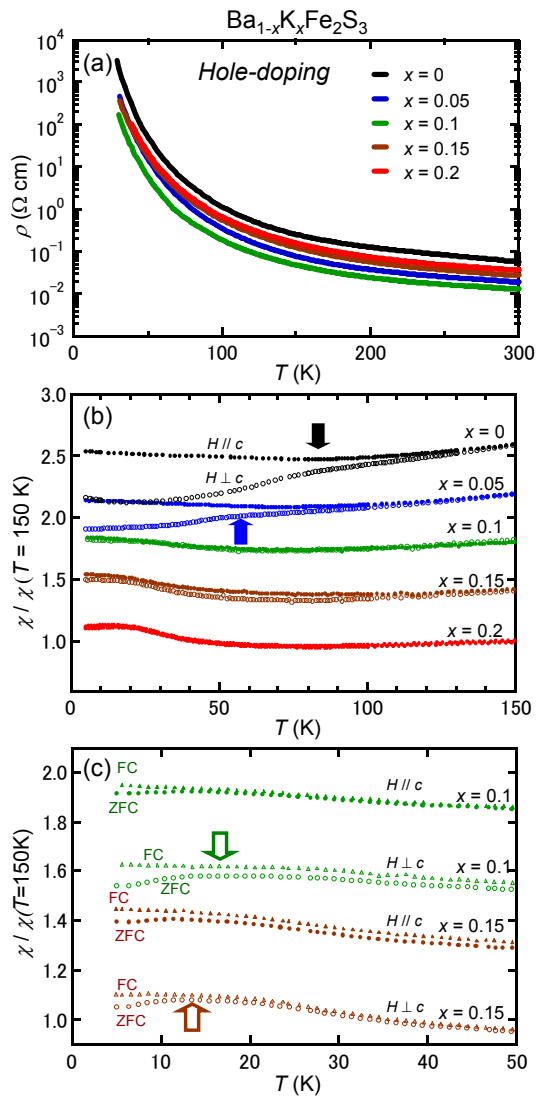}%
\caption{\label{fig:4} (a) Temperature ($T$)-dependence of the resistivity ($\rho$) of Ba$_{1-x}$K$_{x}$Fe$_{2}$S$_{3}$. (b) Temperature ($T$)-dependence of the normalized magnetic susceptibility $\chi$/$\chi(T = 150 {\rm K})$ under the magnetic field of 5 T for Ba$_{1-x}$K$_{x}$Fe$_{2}$S$_{3}$ parallel (solid circle) and perpendicular (open circle) to the $c$-axis. Each curve is shifted by the offset of 0.4. The solid arrows denote the antiferromagnetic transition temperature ($T_{\rm N}$). (c) Extended figure of $\chi$/$\chi$($T = 150$ K) curve in the low temperature region for $x$ = 0.1 and 0.15. Both field cooled (FC) curve (solid/open triangle) and zero field cooled (ZFC) curve (solid/open circle) are shown. The open arrows indicate the spin-glass transition temperature ($T^{*}$.)}
\end{figure}

Figures~\ref{fig:1} (d), \ref{fig:1} (e) show the relations between nominal and actual K/Co contents estimated by SEM/EDX analysis for Ba$_{1-x}$K$_{x}$Fe$_{2}$S$_{3}$ and BaFe$_{2-y}$Co$_{y}$S$_{3}$ . Note that the $\delta$ value is fixed to be 0 in these samples. The actual K/Co content is roughly same as the nominal value, which indicates that K/Co atoms are successfully substituted at $x < 0.2$ and $y < 0.2$. Hereafter, we use nominal $x$ and $y$ values to identify the samples.

Figure \ref{fig:4} (a) shows the $T$-dependence of $\rho$ of Ba$_{1-x}$K$_{x}$Fe$_{2}$S$_{3}$ along the $c$-axis. For all $x$ investigated in this work, even though the hole carriers are expected to be doped to the conduction band, $\rho$ still shows an insulating behavior. The $T$-dependences of $\rho$ are qualitatively described by the 1D VRH conduction. Figure \ref{fig:4} (b) presents the $T$-dependence of $\chi$ for Ba$_{1-x}$K$_{x}$Fe$_{2}$S$_{3}$ under the magnetic field of 5 T along the $c$-axis, and perpendicular to the $c$-axis (a certain intermediate direction between the $a$-axis and $b$-axis), for Ba$_{1-x}$K$_{x}$Fe$_{2}$S$_{3}$. $T_{\rm N}$ can be found at the temperature where the anisotropy of $\chi$ starts to emerge on cooling (indicated by solid arrows in Fig.~\ref{fig:4} (b)). By substituting K atoms into the Ba sites, $T_{\rm N}$ rapidly decreases and the long-range magnetic ordering is indiscernible  at $x >$ 0.1. Instead, at  $x >$ 0.1, we can recognize a discrepancy between the zero-field cooled and field-cooled cycles, suggesting the spin-glass-like magnetism (indicated by open arrows in Fig.~\ref{fig:4} (c)). The spin-glass transition temperature is $T^{*}$ = 17 K at $x$ = 0.1, which decreases with increasing $x$ and seems to vanish at $x$ = 0.2. The spin glass transition is also reported in the Se-analog compound, Ba$_{1-x}$K$_{x}$Fe$_{2}$Se$_{3}$ with much higher $T^{*}$. This is probably due to weaker magnetic interaction in BaFe$_{2}$S$_{3}$ than that in BaFe$_{2}$Se$_{3}$.\cite{caron}

\begin{figure}
\includegraphics[width=7cm]{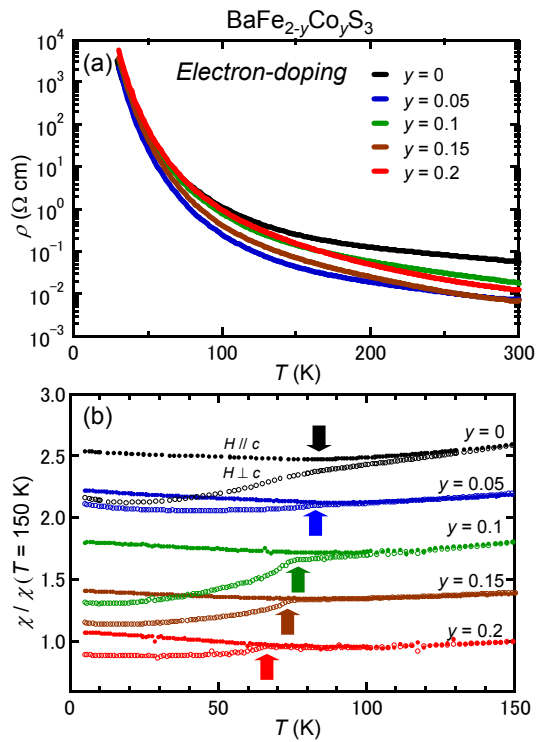}%
\caption{\label{fig:5} (a) Temperature ($T$)-dependence of the resistivity ($\rho$) of BaFe$_{2-y}$Co$_{y}$S$_{3}$. (b) Temperature ($T$)-dependence of the normalized magnetic susceptibility $\chi$/$\chi$($T = 150$ K) of BaFe$_{2-y}$Co$_{y}$S$_{3}$ parallel (solid circle) and perpendicular (open circle) to the $c$-axis. Each curve is shifted by the offset of 0.4. The arrows denote the antiferromagnetic transition temperature ($T_{\rm N}$.)}
\end{figure}

Figure \ref{fig:5} (a) shows the $T$-dependence of $\rho$ for BaFe$_{2-y}$Co$_{y}$S$_{3}$ along the $c$-axis. Just like the case of K substitution, all the curves show the 1D VRH conduction. The $T$-dependences of $\chi$ for BaFe$_{2-y}$Co$_{y}$S$_{3}$ under the magnetic field of 5 T both parallel and perpendicular to the $c$-axis are shown in Fig.~\ref{fig:5} (b). Contrary to K substitution, the suppression of $T_{\rm N}$ with Co substitution is rather mild and $T_{\rm N}$ still remains at 67 K even at $y$ = 0.2.

\begin{figure}
\includegraphics[width=8cm]{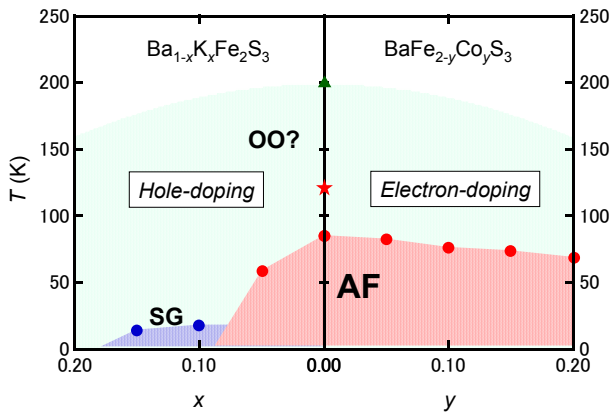}%
\caption{\label{fig:6}The phase diagram of Ba$_{1-x}$K$_{x}$Fe$_{2}$S$_{3}$ and BaFe$_{2-y}$Co$_{y}$S$_{3}$. The red shaded area shows antiferromagnetic (AF) phase. The red solid star indicates the antiferromagnetic transition temperature ($T_{\rm N}$) of nominally off-stoichiometric BaFe$_{2.1}$S$_{3}$. The blue shaded area denotes spin-glass (SG) phase. The green shaded area indicates possible orbital ordering (OO) phase suggested by an anomaly in resistivity curve (indicated by a green triangle).}
\end{figure}

In Fig.~\ref{fig:6}, the results of $\rho$ and $\chi$ measurements for Ba$_{1-x}$K$_{x}$Fe$_{2}$S$_{3}$ and BaFe$_{2-y}$Co$_{y}$S$_{3}$ are summarized as the phase diagram. While K substitution induces 0.5 $x$ holes per one Fe atom, Co-substitution induces 0.5 $y$ electrons per one Fe atom. Therefore, the horizontal axis of Fig.~\ref{fig:6} is proportional to the doping level of electrons. The phase diagram manifests electron-hole asymmetry; K substitution (hole doping) destroys magnetic ordering more effectively than Co substitution (electron doping). The fragile magnetic phase in the hole-doped side in comparison to the electron-doped side is completely opposite to the trend  of 2D iron-based superconductors such as Ba$_{1-x}$K$_{x}$Fe$_2$As$_{2}$ and BaFe$_{2-y}$Co$_{y}$As$_{2}$.\cite{shen}

The explanations of such asymmetry are given in both of itinerant and localized pictures on the electronic states. In the itinerant electron picture, the strip-type magnetism is stabilized by the Fermi surface nesting between hole and  electron pockets. We simply assume that  the mass of electrons is  heavier than that of holes, which is different situations from 2D iron-based superconductors.\cite{coldea,analytis} In this case, moderate electron doping can keep the nesting feature and then sustain the antiferromagnetic state. On the other hand, small amount of hole doping easily weaken the nesting feature, destabilizing the antiferromanegtic state. Thus, the observed electron-hole asymmetry can be reproduced. A recent first-principle calculations for BaFe$_{2}$S$_{3}$ revealed that indeed one of the electron pockets, which is predominantly responsible for magnetic correlation, is quite shallow and can quickly shrink with hole doping.\cite{arita} In this scenario, if there are multiple hole and electron pockets, the most well-nested Fermi surface would be realized under electron doping, so one can generally anticipate that the top of the dome of $T_{\rm N}$ would locate at the finite $y$ value; however, $T_{\rm N}$ is highest exactly at $x$ = 0 in the experimentally obtained phase diagram. The reason is probably the disorder effect on the Fe plane induced by Co substitution, which destabilize antiferromagnetic ordered state.\cite{du_co} The superconductivity under high pressure likely emerges owing to large magnetic fluctuations originating from the Fermi surface nesting; therefore, the fine tuning of doping level can potentially improve the superconducting functionalities such as the critical temperature and the critical pressure.

In the localized electron picture, the electron-hole asymmetry is well accounted for by assuming that the holes are doped not into the Fe 3$d$ orbitals but into the ligand S 3$p$ orbitals in the hole-doped system; this likely happens because the Fe$^{3+}$ ions are high-valence in the chalcogenides and belong to the negative charge-transfer regimes.\cite{atanasov} The antiferromagnetic interaction between Fe 3$d$ electron spins and S 3$p$ hole spin is considered to produce an effective ferromagnetic interaction between two Fe spins locating adjacent to each other across S site, which competes with the antiferromagnetic superexchange interaction between two adjacent Fe spins. This magnetic frustration breaks the antiferromagnetic ordering rapidly. On the other hand, the doped electrons are introduced to the Fe sites, so that there appears no magnetic frustration, resulting in the robust antiferromagnetic phases. These electron-hole asymmetry is the characteristic behavior of a doped Mott insulator, which is quite analogous to the physics in copper based superconductors.\cite{armitage} In this scenario, therefore carrier doping, especially hole doping, may improve the critical temperature and the critical pressure of the superconductivity. In order to determine which scenario is more plausible, further theoretical and experimental studies on electronic and magnetic structures are required.

\section{Conclusion}
We have investigated the effect of stoichiometry and substitution on the electronic and magnetic properties of iron-based ladder compound BaFe$_{2}$S$_{3}$. The resistivity and magnetization are quite sensitive to the off-stoichiometry, and the antiferromagnetic transition temperature reaches highest value 122 K in true stoichiometric composition. The antifferomagnetic ordering is robust against Co doping, while it is rather easily broken by K doping; such electron-hole asymmetry can be understood based on both itinerant and localized electron pictures.

We are grateful to H. Takahashi, F. Du, T. Hawai, Y. Ueda, and R. Arita for helpful discussions and experimental support. This research was supported by the Grant Program of the Yamada Science Foundation and MEXT Elements Strategy Initiative to Form Core Research Center.

\end{document}